# APPLICATION OF SCANNING MID-IR-LASER MICROSCOPY FOR CHARACTERIZATION OF SEMICONDUCTOR MATERIALS FOR PHOTOVOLTAICS

**V. P. Kalinushkin [†], O. V. Astafiev [‡ i], and V. A. Yuryev [‡]**

*[†] General Physics Institute of RAS, 38 Vavilov Street, Moscow 117942, Russia.*

*Tel.: +7 (095) 132 8126. Facsimile: +7 (095) 135 1330. E-mail: vkalin@kapella.gpi.ru.*

*[‡] Natural Science Center of General Physics Institute of RAS.*

*Tel.: +7 (095) 132 8144. Facsimile: +7 (095) 135 1330. E-mail: vyuryev@kapella.gpi.ru.*

## Abstract

The scanning mid-IR-laser microscopy was previously demonstrated as an effective tool for characterization of different semiconductor crystals. Now the technique has been successfully applied for the investigation of CZ $Si_xGe_{1-x}$ — a promising material for photovoltaics — and multicrystalline silicon for solar cells. In addition, this technique was shown to be appropriate for imaging of polishing-induced defects as well as such huge defects as "pin holes". Besides, previously unexplained "anomalous" (cubic power) dependence of signal of the scanning mid-IR-laser microscope in the optical-beam-induced light scattering mode on the photoexcitation power obtained for mechanically polished samples has now been attributed to the excess carrier scattering on charged linear defects, likely dislocation lines. The conclusion is made in the article that the scanning mid-IR-laser microscopy may serve as very effective tool for defect investigations in materials for modern photovoltaics.

[i] Present address: Tokyo University, Department of Basic Science, Komaba 3-8-1, Meguro-ku, Tokyo, 153. E-mail: astf@mujin.c.u-tokyo.ac.jp.

The scanning mid-IR-laser microscopy was previously demonstrated as an effective tool for characterization of different semiconductor crystals. [1–4] Now it is introduced as a tool for characterization of materials for modern photovoltaics.

### 1. The instrument brief description.

A pictorial diagram of the scanning mid-IR-laser microscope is given in Fig. 1. Presently, the instrument operates in two main modes: a mode of the scanning low-angle mid-IR light scattering (SLALS) — the basic regime of its operation — which is sensitive to aggregations of ionized point defects (or free carrier accumulations) and any inhomogeneities in the distribution of free carriers (some restrictions are imposed on the accumulation or inhomogeneity characteristic sizes and its profile function,[3] however) and a mode of the optical beam induced light scattering (OLALS) — a direct optical analog of the electron-beam-induced current (EBIC) or optical-beam-induced current (OBIC) methods — which reveals recombination active defects and enables carrier lifetime mapping. Both modes are the dark-field ones. Besides, analogous regimes of the microscope operation are available in the bright field (mid-IR light absorption) sub-mode.[ii] (A relatively detailed description of the technique as well as the method of the low-angle mid-IR-light scattering, on the basis of which the scanning mid-IR-laser microscopy was developed, illustrated with main results and containing rather complete citation is given in Ref. 3. In addition, a detailed description of the low-angle mid-IR-light scattering technique with many results and references can be found in a large but rather old article cited in Ref. 5.)

At present, low-temperature SLALS and OLALS facilities which would enable the defect composition analysis[3] as well as the magneto-optical SLALS are under the develop-

---

[ii] Mention that all the results given in the current article were obtained using $CO_2$-laser as a probe radiation source ($\lambda$=10.6 μm).

ment. The defect profile option and quantitative carrier lifetime and concentration mapping will also be available soon.

## 2. Defect images and experimental dependences.

### 2.1. Defects in CZ Si$_{1-x}$Ge$_x$.

SLALS and OLALS micrographs of the samples of single-crystalline CZ Si$_{1-x}$Ge$_x$ alloy with Ge content from 2.2 to 4.7 at. % are presented in Fig. 2. Two areas were revealed in the X-ray topographs of these crystals: the area free of striation and dislocations around the wafer centers (area I) and the area containing striation and dislocations in the periphery of the wafers (area II).[4] SLALS pictures shows the striation in the area II and no striation in the area I. OLALS pictures demonstrate that no or low (Fig. 2 (*h*)) recombination contrast is usually caused with the grown-in striations in the crystal bulk, although a high contrast was revealed in Fig. 2 (*l*). The second type of defects manifested as dark stripes in the OLALS micrographs (Fig. 2 (*b*),(*l*)) can likely be identified as dislocations and dislocation walls which are registered in X-ray patterns of the area II and revealed by etching. The last type of defects observed are those seen as black spots in the OLALS patterns (Fig. 2 (*b*),(*d*),(*f*),(*h*),(*j*)). They were present in both areas and appeared to have a non-dislocation origin: some non-dislocation defects found in both areas by the selective etching may be similar to the defects revealed by OLALS. The latter defects seem to be the main lifetime (and cell efficiency) killing extended defects in the studied material.

### 2.2. Visualization of grain boundaries in multicrystalline Si.

Fig. 3 shows the recombination contrast images of grain boundaries in two samples of multicrystalline silicon for solar cells (the optical-beam-induced absorption sub-mode). It is clear that scanning mid-IR-laser microscopy can be effectively used for the investigation and testing of the efficiency of grain boundary passivation in this material.

### 2.3. Visualization of polishing-induced defects in single crystalline Si.

Fig. 4 demonstrates the OLALS images of areas on single-crystalline CZ Si wafers subjected to different processing procedures. In all the pictures, the recombination-active defects (dark regions) are clearly seen. A chaotic pattern of defects in the wafer subsurface region which arise after the mechanical polishing (Fig. 4 (*a*)) transforms in the stripe-like one due to the mechano-chemical polishing (Fig. 4 (*b*)). The latter defects seem to be the traces of underpolished scratches (remark that scratches have been revealed on this wafer neither by visual inspection using optical microscope nor by inspection using electron microscope). Moreover, these defects supposingly give rise to some defective areas revealed in the subinterface region of the wafers subjected to the oxidation procedure (Fig. 4 (*c*)).

The above example gives an evidence that the scanning mid-IR-laser microscopy can serve as an efficient tool for testing the wafer treatment quality.

### 2.4. "Anomalous" dependence of the microscope signal on the photoexcitation power in the optical-beam-induced scattering mode obtained for mechanically polished Si samples.

The dependences of the detector signal on the photoexcitation power were reported in Ref. 2 (see Fig. 5) for both optical-beam-induced scattering and absorption sub-modes of the scanning mid-IR-laser microscope for mechanically and mechano-chemically polished sides of the same single-crystalline silicon wafer. Usual for linear recombination second-power law[3,5] was found in the optical-beam-induced scattering sub-mode for mechano-chemically polished side. In the induced absorption sub-mode, a linear law was obtained for both wafer sides which is also in agreement with the theoretical prediction. Only for the mechanically polished side an "anomalous" cubic dependence was registered in the induced light-scattering sub-mode which has not been explained thus far.[iii] Now the explanation has been found.

Let us suppose that charged linear defects (dislocation lines) play the main role in the excess carrier scattering process and the defect density $N_m$ (per unit area) in the investigated subsurface layer damaged with the mechanical polishing is so high that $\omega << 1/\tau$. Following Ref. 6 it can be obtained for charged linear defects that

$$\mathrm{Im}(\varepsilon) \approx \frac{2^{9/2} n^{3/2} \varepsilon_{\mathrm{L}}^{3/2} \varepsilon_0^{1/2} e\, kT}{\pi^{1/2} N_{\mathrm{m}} q^2 m^{*1/2} \omega\, g(p)} \tag{1}$$

where

$$g(p) = (1 + 2p)\exp(p)\,\mathrm{erfc}\left(p^{1/2}\right) - 2\,p^{1/2}/\pi^{1/2}, \tag{2}$$

$$p = \frac{\hbar^2 e^2 n}{8\, m^* \varepsilon_{\mathrm{L}} \varepsilon_0 (kT)^2} \tag{3}$$

In Eq. (1)–(3), $\tau$ is the momentume relaxation time, $n$ is the free carrier concentration; $\varepsilon_{\mathrm{L}}$ is the lattice dielectric constant; $q$ is charge per defect unit length; $m^*$ is the carrier effective mass; $\omega$ is the probe light cyclic frequency; the rest designations are used in their commonly known meaning. The equations are given in SI units.

For the light scattering $I_{sig} \sim I_{sc} \sim \left|\delta\varepsilon\right|^2 \sim \left[\mathrm{Im}\,\varepsilon\right]^2$.[3,5]

Hence it can be concluded from Eq. (1)–(3) assuming the linear recombination that

$$I_{sig} \sim n^3 \sim W_{ex}^3. \tag{4}$$

Thus, the discussed "anomalous" dependences of the light scattering intensity (detector signal) on the photoexcitation power obtained for the mechanically polished wafers (as

---

well as the cubic law previously obtained for Ge) can be satisfactorily explained supposing that the subsurface layer in such wafers contains a great number of dislocations which dominate in the process of the free carrier scattering.

### 2.5. Superlarge defects in CZ Si:B.

Fig. 6 demonstrates the images of superlarge defects previously observed in CZ Si:B by means of EBIC and the law-angle mid-IR-light scattering technique (see Ref. 7). These defects are met very rare in the industrial Si wafers, nevertheless they can be found sometimes. It is clear that such huge and active defects may destructively affect the efficiency of solar cells. We suppose now taking in the account the size, shape and "strength" of these defects that they are so-called "pin holes" known in the single-crystalline silicon technology.

### 3. Conclusion.

Summarizing the above we can conclude that the technique of the scanning mid-IR-laser microscopy can be considered as a convenient, easy to use but very effective tool for materials studies in the branch of photovoltaics. We have illustrated the technique with only a few examples of results obtained by means of it. It can be used, however, for solving a great number of physical and technological problems. Evidently, it could be successfully used for solving the problems of wafer surface pollution, washing, *etc.*, and the effect of them on the surface recombination velocity, like it was done in Ref. 8 by means of a technique which was a very close analog of the induced absorption sub-mode of the presently described instrument. In addition, the problems related to the precipitated and interstitial oxygen distribution in silicon might be solved by means of the described technique, practically like it was done in Ref. 9. The advantage of application of the scanning mid-IR-laser microscopy to solving such class of problems is its higher sensitivity due to the dark field compared to the IR-light absorption-based methods.

An additional undoubted merit of the method of the mid-IR-laser microscopy is that it is relatively inexpensive that is important for the industrial practice.

### Acknowledgments.

The authors express their appreciation to the Ministry of Science and Technologies of the Russian Federation for the financial support of this work which is carried out within the framework of the sub-program "Perspective Technologies and Devices of Micro- and Nanoelectronics" under the grant No. 02.04.3.2.40.Э.24.

**Fig. 1.** An optical diagram of the scanning mid-IR-laser microscope: (1) the probe IR-laser beam (routinely, CO or $CO_2$-laser), (2) the studied sample (can be placed in a cryostat or a furnace) in the front focal plane of the lens L1, (3) an aperture with a diameter $D_1$ in the plane of the lens L1, (4) the lens L1, (5) an opaque screen or a mirror with a radius $b_1$ in the back focal plane of the lens L1, (6) a diaphragm with a radius $b_0$ in the back focal plane of the lens L1, (7) an aperture with a diameter $D_2$ in the plane of the lens L2, (8) the lens L2, (9) the scattered wave, (10) an IR photodetector in the back focal plane of the lens L2, (11) an exciting laser beam (used in the OLALS—recombination contrast—mode).

**Fig. 2.** Couples of SLALS (*a*),(*c*),(*e*),(*g*),(*i*),(*k*) and OLALS (*b*),(*d*),(*f*),(*h*),(*j*),(*l*) micrographs of the same regions of $Si_{1-x}Ge_x$ wafers (1×1 mm); (*a*)–(*d*): p-type CZ Si (100), 4 at. % of Ge; (*e*)–(*h*): p-type CZ Si (111), 4.7 at. % of Ge; (*i*)–(*l*): n-type CZ Si (111), 2.2 at. % of Ge; (*e*),(*f*),(*i*),(*j*) are close to the wafer centers, the rest are far from the centers.

**Fig. 3.** Micrographs of multicrystalline silicon for solar cells (the optical-beam-induced absorption submode, 4×4 mm). The darker the image the shorter the lifetime is. Grain boundaries are clearly seen in the images.

**Fig. 4.** OLALS micrographs of Si wafers (2×2 mm); (*a*): mechanical polishing; (*b*): mechano-chemical polishing (likely underpolished wafer); (*c*): under $SiO_2$ layer (1200 Å thick); polishing-induced defects are seen in all the pictures as black areas.

**Fig. 5.** OLALS: dependence of the detector signal on the photoexcitation power for mechanically (1),(3) and mechano-chemically (2),(4) polished sides of the same CZ Si:B wafer;

(1),(2): optical-beam-induced light scattering (dark field) submode; (3),(4): optical-beam-induced absorption submode; (1): $I_{sig} \sim W_{ex}^3$ ; (2): $I_{sig} \sim W_{ex}^2$ ; (3),(4): $I_{sig} \sim W_{ex}$ .

**Fig. 6.** Superlarge defects in CZ Si:B: couple of SLALS (*a*) and OLALS (*b*) micrographs of the same region of CZ Si:B wafer (1×1 mm) and EBIC micrograph of analogous defects in a different wafer of the same material (*c*).

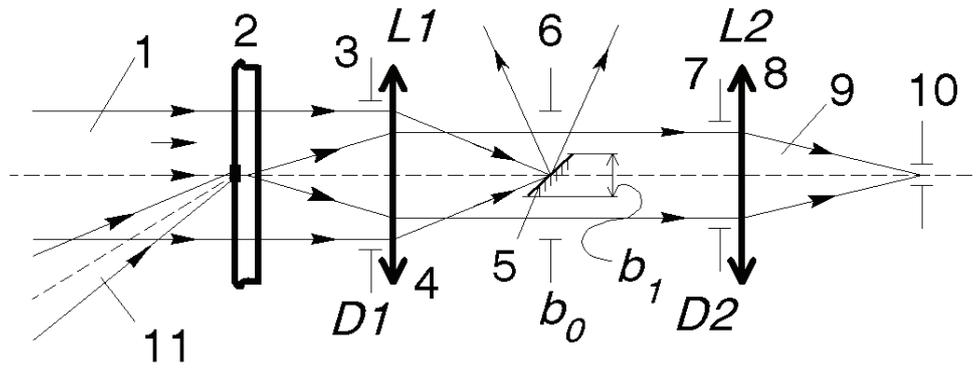

Fig. 1.

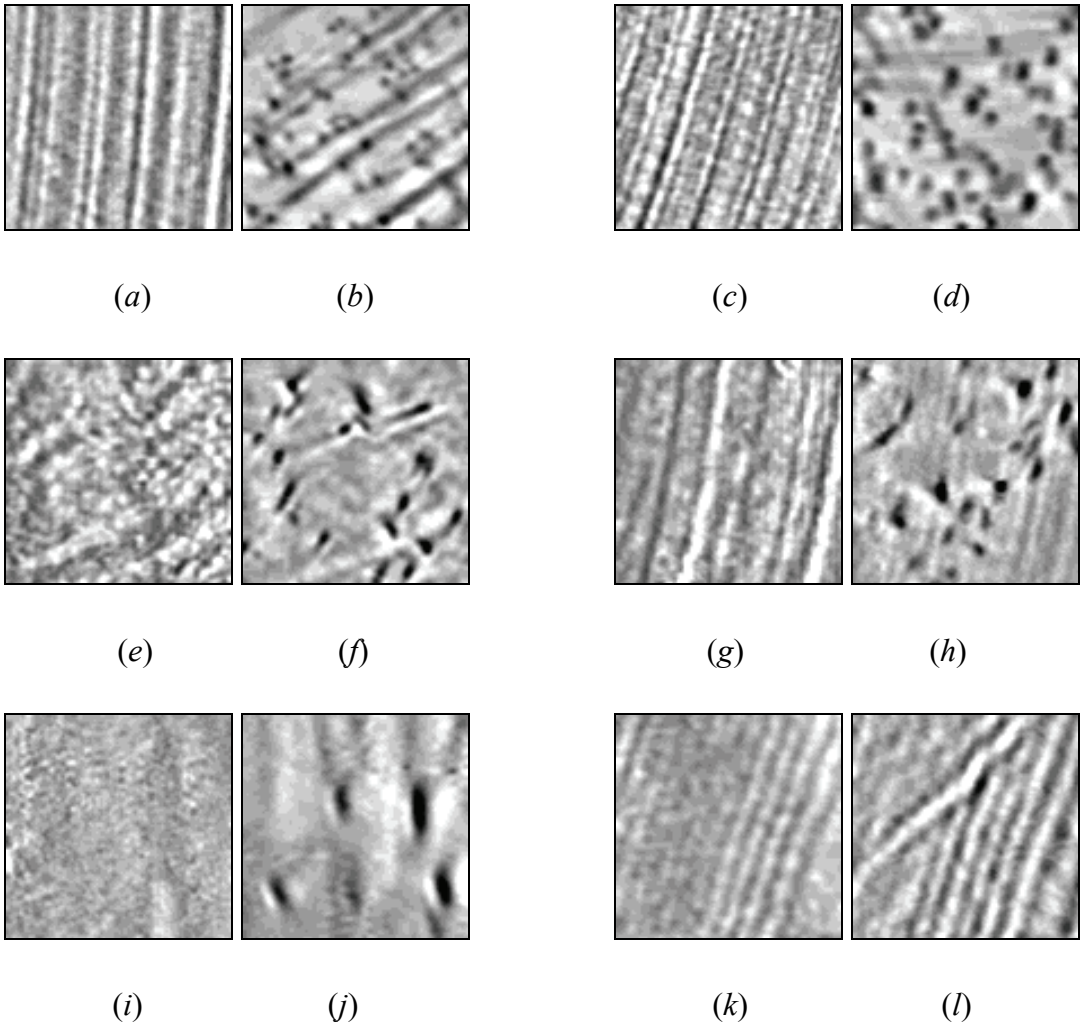

(a)      (b)         (c)      (d)

(e)      (f)         (g)      (h)

(i)      (j)         (k)      (l)

Fig. 2.

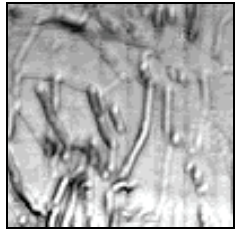 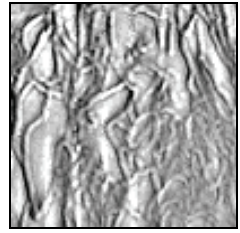

(*a*)        (*b*)

Fig. 3.

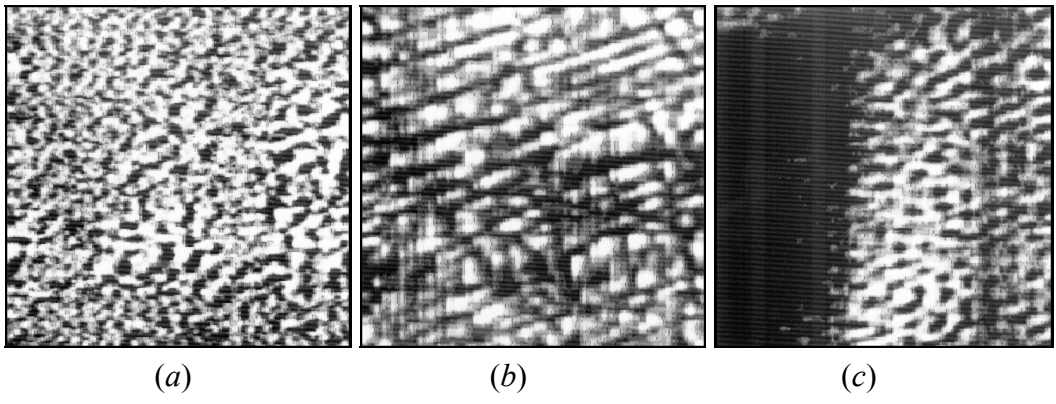

(*a*)        (*b*)        (*c*)

Fig. 4.

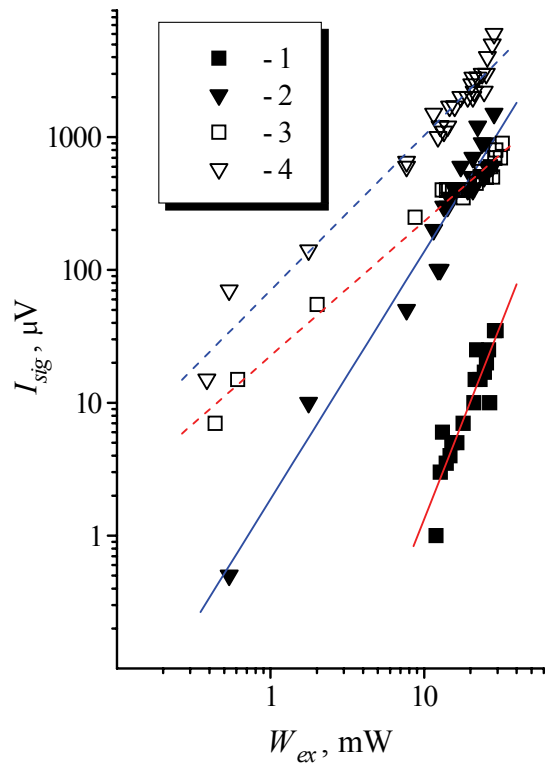

Fig. 5.

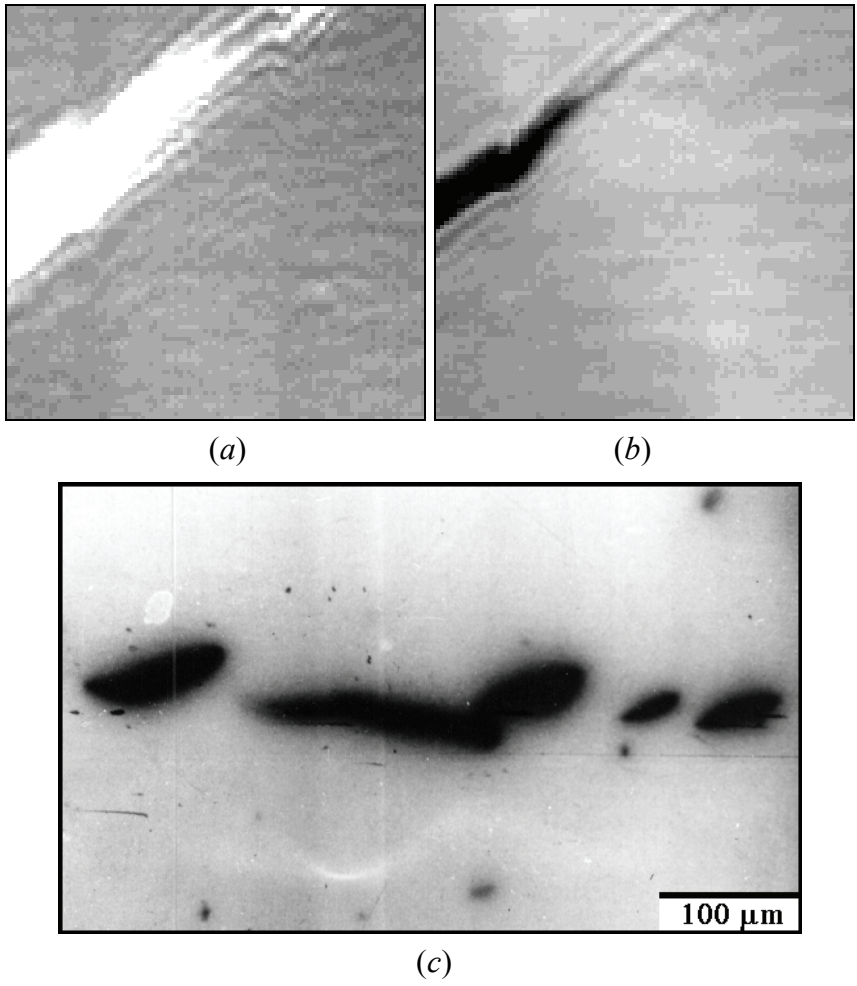

(a)          (b)

(c)

Fig. 6.